\documentstyle[aps,prl,preprint]{revtex}
\begin{document}
\title{Phases of the 2D Hubbard model at low doping}
\author{Andrey V. Chubukov \and Karen A. Musaelian}
\address{Department of Physics, University of Wisconsin - Madison,
1150 University Avenue, Madison, WI 53706}
\date{\today}
\maketitle
\begin{abstract}
We show that the planar spiral phase of the 2D Hubbard model
at low doping,$x$, is unstable towards a noncoplanar spin configuration.
The novel equilibrium state we found at low doping
is incommensurate with the inverse pitch of the spiral varying as $\sqrt{x}$,
but nevertheless has a dominant peak in the susceptibility at $(\pi,\pi)$.
Relevance to the NMR and neutron scattering experiments in
$La_{2-x}Sr_{x}CuO_4$ is discussed.
\end{abstract}
\narrowtext
Magnetic properties of the $CuO_2$ layers in the
high temperature superconductors have been recently
attracting an intense interest~\cite{magnetism}
as magnetism is possibly a major contributor to
 the  mechanism of superconductivity.
Among cuprate superconductors, $La_2CuO_4$
 holds a special place as the most simple and extensively experimentally
studied
 compound. There are numerous reasons to believe that magnetic
 properties of $La_{2-x}Sr_xCuO_4$ are quantitatively
 captured by the 2D Hubbard model with
chiefly the nearest-neighbor hopping
\begin{equation}
{\cal H} = -t \sum_{i,j} a^{\dagger}_{i,\alpha} a_{j,\alpha} + U \sum_{i}
n_{i,\uparrow} n_{i,\downarrow}
\label{hamilt}
\end{equation}
Here $\alpha$ is a spin index, and $n = a^{\dagger} a$.
At half-filling,
the ground state of the 2D
Hubbard model exhibits a long range commensurate
N\'{e}el order.
However, the  introduction of even a very
 small number of holes into the system
brings about a dramatic change in the nature of the ground state as
the $T=0$ static Pauli susceptibility associated with vacancies
does not depend on the carrier concentration
in two spatial dimensions.

Shraiman and Siggia pointed out~\cite{siggia} that holes introduced
into a commensurate antiferromagnet give rise to a long-range dipolar
distortion of the staggered magnetization. In the simplest scenario,
this leads to a spiral spin
configuration with the momentum $(\pi,Q)$. The
incommensurate $(\pi,Q)$ spin-density-wave phase was also obtained in the early
perturbative studies of the Hubbard model with small $U$~\cite{Schultz}
 and in several other
mean-field~\cite{john,lee} and self-consistent~\cite{subir} calculations.
 Experimental situation in
$La_{2-x}Sr_{x}CuO_{4}$ is  however far from being clear. On one hand,
the inelasic neutron scattering experiments~\cite{incommen}
have shown that  $\chi^{\prime \prime}$ is peaked
at incommensurate $(\pi,Q)$ and symmetry related wave vectors,
 with $Q$ varying as $\pi-Q \approx 2 \pi x$.
On the other hand, the longitudinal spin-lattice
relaxation rate data for $^{17}O$~\cite{nmr}, obtained at far smaller
frequencies
cannot be fitted by the form of $\chi^{\prime \prime}$ inferred
from the neutron scattering experiments.

In this communication we use the RPA formalism of Schrieffer, Wen and
Zhang~\cite{SWZ} and study in detail the
structure of magnetic correlations in the Hubbard model at small
doping. We will show on the
basis of an analysis of  low frequency bosonic modes
that the $(\pi,Q)$ spiral state is actually
{\em unstable} with respect to a more complex {\em noncoplanar} spin
configuration which, though incommensurate, has a dominant peak in the
dynamical
susceptibility precisely at $(\pi,\pi)$.

The analysis presented here is related to other works on incommensurate
magnetic phases at finite doping. Previous mean-field studies of the
Hubbard and $t-J$ models~\cite{siggia,john,lee} have shown that
in a certain range of parameters, the $(\pi,Q)$ spiral
state has lower energy than both the
N\'{e}el state and the spiral state with the pitch
in both spatial directions.
Our energy analysis  is consistent with their results.
Shraiman and Siggia~\cite{boris}
 developed a macroscopic theory of the bosonic
excitations in the
spiral phase. To the lowest order in density,
they found a peculiar degeneracy in
 the ground state energy
for the planar spiral state
and for a whole set of
noncoplanar magnetic configurations with the plane of the spiral varying in
space. This degeneracy is also present in our  microscopic calculations.
 However, the further assumption of Shraiman and Siggia that
the next-order terms in doping concentration  stabilize
the planar spiral
state is inconsistent with our microscopic results for the Hubbard model.

As an input for our analysis, we will
need an expression for the energy spectrum, $E_k$,
 of a single hole in a quantum antiferromagnet.
The mean-field theory gives
 $E_k= -~\sqrt{\epsilon_k^2+\Delta^2}$, where $\epsilon_k=
-2t(cosk_x+cosk_y)$ and $\Delta$ is a gap which separates valence and
conduction bands~\cite{john,chub_fren,SWZ,tes}.
This energy is obviously degenerate along the whole edge
of the magnetic Brillouin zone
$\vert k_x\pm k_y \vert = \pi$, where $E_k$ has a maximum.
However, both perturbative~\cite{vignale} and
variational~\cite{Manous} studies have shown that
this degeneracy does not survive the effects of quantum fluctuations, and the
actual dispersion has a maximum at
four points $(\pm\pi/2,\pm\pi/2)$
in the center of each of the edges of the magnetic Brillouin zone. In the
neighborhood of these points $E_k$ can be presented as
$E_k=-\Delta + p_{\parallel}^{2}/2m_{\parallel} + p_{\perp}^{2}/2m_{\perp}$,
where near $(\pi/2,\pi/2)$,
 $p_{\parallel}=(k_x - k_y)/2, ~p_{\perp}=(k_x + k_y)/2$.
Self-consistent calculations predict that
at large $U$ both masses scale as inverse bandwidth $J =4t^2/U$, but
numerically, $m_{\parallel}$ (which is infinite in the mean-field theory)
is several times larger than $m_{\perp}$.

We now turn to the description of our calculations.
Consider first
a N\'{e}el ordered state at small but finite doping. Near ${\bf Q}_0 =
(\pi,\pi)$,
the static transverse
susceptibility should have a hydrodynamic form~\cite{hydro}
$~\chi^{xx}_{\rm st}(\bf q \approx \bf Q_0) =
 N_0^2 /(\rho_s(\bf q-\bf Q_0)^2)$,
where ${\bf N}_0$ is the on-site magnetization, and $\rho_s$ is the
spin-stiffness.
 This form of transverse susceptibility
is reproduced in the RPA formalism by summing up the ladder series of bubble
diagrams.
At half-filling, only bubbles containing valence and conduction fermions are
allowed, while at finite doping chemical potential moves inside the valence
band, and one also has contributions to $\chi$
from  bubbles with only valence
fermions. These last contributions are proportional to 2D Pauli susceptibility,
which, as we said, does not depend on carrier concentration. As a result, one
obtains a finite correction to the spin stiffness already at a very
small doping~\cite{chub_fren}:
\begin{equation}
\rho_s = \rho_s^0 (1-z),~~z = 4T\chi_{\rm 2D}^{\rm pauli} =
\frac{2T\sqrt{m_{\perp}m_{\parallel}}}{\pi}
\label{stiff}
\end{equation}
Here $\rho_s^{0}$ is the spin stiffness at half filling, and
 $T$ is the scattering amplitude for two holes.
The latter is equal to $U$ in
the mean-field description, but strong
self-energy and vertex corrections  reduce $T$ to the order of the bandwidth
$J$~
\cite{siggia,chub_fren,klr}, which in turn implies that
 $z$ is simply a dimensionless number.
For the rest of the paper, we will consider
$z$ as a phenomenological input parameter~\cite{comment}.
It follows from (\ref{stiff}) that N\'{e}el state remains stable at finite
doping if
$z <1$, but becomes unstable if $z>1$, which implies that we have to
consider incommensurate spin configurations as
possible candidates for the ground states.

Let us first focus on the two simplest candidates \cite{siggia,subir,piers}:
 the spiral states with the ordering vectors $(\pi, Q)$ and $(Q,Q)$.
For definiteness, we choose the ordering to be in the XY plane such that
 $S^X_R = S_{\bar Q}cos({\bar {\bf Q}}{\bf R})$, and $S^Y_R = S_{\bar
Q}sin({\bar {\bf Q}}{\bf R})$,
where ${\bar {\bf Q}}$ is the ordering momentum.

The mean-field analysis for the incommensurate states proceed in the same way
as for the N\'{e}el state. We skip the details of the calculations
and focus only on the results. For $(\pi,Q)$ state, we obtained
$Q = \pi -(U/t) x$ and
\begin{equation}
E^{(\pi,Q)}-E^{(\pi,\pi)}= \frac{\pi x^2}{4\sqrt{m_\perp m_\parallel}}(1-z)
\label{epiq}
\end{equation}
 Clearly, the $(\pi,Q)$ phase has lower energy than the $(\pi,\pi)$ phase
for $z>1$, exactly where the susceptibility of the $(\pi,\pi)$ state becomes
negative.

For $(Q,Q)$ phase, the inverse pitch $Q$
 is the same, but the energy difference is
\begin{equation}
E^{(Q,Q)} - E^{(\pi,\pi)} = \frac{\pi x^2}{4\sqrt{m_\perp m_\parallel}}
(3-2z)~.
\label{eqq}
\end{equation}
Compairing (\ref{epiq}) and (\ref{eqq}), we observe that
the $(\pi,Q)$ spiral phase has the lowest energy
at $1<z<2$. This is roughly consistent with the results of other mean-field
approaches~\cite{siggia,lee}. At $z>2$, the $(Q,Q)$ state
has the lowest energy. However, we found that $\partial E^{(Q,Q)}/\partial x^2$
is negative at $z>2$. This suggests phase
separation which is probably
unphysical because the model neglects long-range Coulomb
interaction~\cite{siggia}. In
view of this, we will only consider the case $1<z<2$.

We now turn to the central topic of our paper, which is
the study of collective bosonic excitations (poles of dynamical susceptibility)
in the $(\pi,Q)$ spiral phase.
The $SO(3)$ symmetry of spin rotations is completely broken in the spiral
phase, and one, therefore, should expect to have
three bosonic zero modes with two different spin-wave velocities.
The calculations of the susceptibilities
is straightforward but lengthly because we have to solve a set of four coupled
Dyson equations. For in-plane $(XY)$ spin fluctuations, a
simple symmetry analysis
predicts that the zero modes are located at ${\bf q} =\mp {\bar {\bf Q}}$.
 for $\chi^{+-}_q$ and $\chi^{-+}_q$ correspondingly.
We indeed found that at $\tilde {\bf q} = ({\bf q} + {\bar {\bf Q}}) \ll 1$,
 static $\chi^{+-}_q$ has the form:
$\chi^{+-}_q = 2/(J{\tilde {\bf q}}^2 (2-z))$, i.e., it is {\it positive}
for $z <2$. For the dynamical susceptibility, we found a pole at
$\omega = c \tilde q$ with the same $c$ (up to $O(x)$ terms) as at
half filling. This is in agreement with other results~\cite{boris,piers}.

We now turn to the magnetic susceptibility $\chi^{zz}_{q}$
associated with the fluctuations of the plane of spin ordering.
 This channel is coupled to
the charge and in-plane spin fluctuations only dynamically, so
for the full static susceptibility
one has the simple RPA formula
$\chi_q^{zz}(\omega=0) = \overline\chi^{zz}_q(\omega = 0)/(1
-U\overline\chi^{zz}_q(\omega = 0))$ where $\overline\chi^{zz}$ is the bare
bubble. From general symmetry considerations, we would expect
the Goldstone modes in $\chi^{zz}_q$ to be at  ${\bf q}=\pm {\bar {\bf Q}}$.
Performing calculations to the lowest nontrivial order in
the density (i.e., to $O(x^2)$), we have,
indeed, found that
$(\chi_q^{zz})^{-1}$ at ${\bf q}=\pm {\bar {\bf Q}}$ is equal to zero. However,
with the same accuracy, we also found that the stiffness for
excitations near these momentum is equal to zero, which in turn
 means that to order $O(x^2)$, all fluctuation
modes between ${\bar {\bf Q}}$ and $-{\bar {\bf Q}}$, including the mode at
$(\pi,\pi)$, remain
degenerate and gapless.
A similar degeneracy was found in the macroscopic consideration by Shraiman and
Siggia~\cite{boris}.
They further
asssumed that $O(x^3)$ terms make the spin
stiffness at $\pm {\bar {\bf Q}}$ positive. This scenario yields
anomalously small spin-wave velocity for out-of-plane fluctuations,
 $v \propto \sqrt{x}$, susceptibility at $(\pi,\pi) \propto x^{-3}$,
 and the doping-dependent quantum corrections to the order parameter
which scale as $\sqrt{x}$ instead of $x$.

We, however, calculated  the static susceptibility
$\chi^{zz}$ at $ {\bf q} = (\pi, \pi-\tilde{q})$
 explicitly to the order $x^3$ and found after tedious algebra:
\begin{equation}
(\chi^{zz})^{-1}_{q} = \frac{4x^3}{U}
\left(1-\frac{2}{z}\right)\left(1 -\frac{\tilde{q}^2}{\bar{q}^2}\right)
\label{instab}
\end{equation}
where ${\bar q} = \pi -Q$. We see that while the
Goldstone mode at $\tilde{q} = \pm \bar{q}$ survives to order $O(x^3)$ (as it
should), the static susceptibility for out-of-plane fluctuations at and near
$(\pi,\pi)$ is {\em negative} for
$z<2$, where mean-field solution favors $(\pi,Q)$ phase. This
implies that the spiral $(\pi,Q)$ state is unstable at low doping. Note that
all $O(x^3)$ terms come from the integration within the hole pockets
(the bubbles with conduction and valence fermions yield regular corrections
in powers of $x^2$). Near the minima, the hole spectrum  has a quadratic
dispersion for an arbitrary form of the hopping term, and we therefore expect
eqn.
(\ref{instab}) to be valid also for the models with further-neighbor hopping,
etc.

We now address the issue of the true ground state at finite doping.
The instability  in $\chi^{zz}$  at
$(\pi,\pi)$, implies that the system prefers to have a
spontaneous commensurate antiferromagnetic order along
$Z$ direction in addition to the incommensurate spin ordering in the XY plane.
 This gives a set of magnetic states which are all noncoplanar (and therefore
have nonzero chirality),
 and differ in the ratio between the
order parameter amplitudes along $Z$ direction
and in the XY plane, Fig.\ref{nonplanar}. Let us first calculate the ground
state energies for this set of states. Clearly,
we have to introduce two gaps $\Delta_{\perp} =
 U\langle S_\perp\rangle $, and
$\Delta_{\parallel} = U\langle S_\parallel\rangle $, where
 $\langle S_\perp \rangle $ and
$\langle S_\parallel \rangle $ are the magnitudes of the off-plane and in-plane
components of the order parameter, respectively.
Performing the mean-field decoupling and the diagonalization
 of the Hubbard
Hamiltonian, we obtained after some algebra
\begin{equation}
E_{\Delta_\perp} = E_{\Delta_\perp \to 0} - Ux^3\left(
\frac{\Delta_\perp}{\Delta_\parallel}\right)^2 \left[\frac{2}{z}-1\right]
\label{nonplan_ener}
\end{equation}
where $E_{\Delta_\perp \to 0}$ is the ground state energy in the limit when
the $Z$-component of order parameter tends to zero,
and the total order parameter
$\Delta^2 = \Delta_\perp^2 + \Delta_\parallel^2$
is fixed by the self-consistency conditions, $\Delta \approx U/2$~\cite{comm2}.
 It is apparent from (\ref{nonplan_ener}) that the energy {\em decreases} as
the ratio $\Delta_{\perp}/\Delta_{\parallel}$ increases. Notice that
(i) the energy dependence on the ratio of $\Delta$ is
in the order $O(x^3)$, while to order $x^2$,
 all states from the set are degenerate
in energy, and (ii) the r.h.s. of (\ref{nonplan_ener})
contains the same positive factor $(2/z) -1$
as the expression for $\chi^{zz}$ at $(\pi,\pi)$.
The inverse pitch of the spiral is related to
$x$ by the requirement that the two self-consistency conditions for two order
parameters be compatible with each other. We obtained (${\bar q} = \pi -Q$)
\begin{equation}
{\bar q} = \frac{U}{t}\frac{\Delta}{\Delta_\parallel}x\left[1+2x\left(
1-\frac{\Delta^2+\Delta_{\parallel}^2}{2\Delta_{\parallel}^2}
\frac{2}{z}\right)\right].
\label{exprq}
\end{equation}
As written, Eqn. (\ref{nonplan_ener}) is valid when ${\bar q}
 \ll \Delta_\parallel/\Delta$. Since ${\bar q}\Delta_\parallel \sim x$
(see (\ref{exprq})), we can keep lowering the energy by  decreasing
 $\Delta_\parallel$
as long as $\Delta_{\parallel} \gg \sqrt{x}$. When both
$\Delta_{\parallel}$ and ${\bar q}$  become of the order of
$\sqrt{x}$, $\Delta E = E_{\Delta_\perp} -E_{\Delta_\perp \to 0}$
 has a more complicated form:
\begin{eqnarray}
\Delta E = \frac{Ux^2}{2}\left[1+(\alpha^2 +2)\beta^2-
\frac{\beta z}{6}\left(\left(\alpha^2+
\frac{8}{z}\right)^{3/2} - \alpha^3\right)\right]\nonumber
\end{eqnarray}
Here we introduced $\Delta_\parallel = \alpha\Delta x^{1/2},~
{\bar q} = \beta (U/t) x^{1/2}$. Eqn (\ref{exprq}) relates $\alpha$ and
$\beta$ as: $\beta = z((\sqrt{\alpha^2+ 4/z}-\alpha)/2$.
Minimazing now the
energy with respect to $\alpha$ we obtain to the lowest order in $2-z$:
$\alpha \approx \sqrt{7 z/(12 (2-z))},~\beta \approx 1/\alpha$, and
\begin{equation}
\Delta E = - \frac{Ux^2}{2}\left[\frac{12}{7}
\left(1-\frac{2}{z}\right)^2\right]
\end{equation}
As expected, $\Delta E$ is negative. Observe also
that it scales as $x^2$, instead of $x^3$ as
in (\ref{nonplan_ener}). Clearly then, the equilibrium state with
$\Delta_\parallel \sim \sqrt{x}$ does not belong,
strictly speaking, to a set of initially degenerate spin configurations, and
could be selected already in the calculations to order $O(x^2)$. The discovery
of the degenerate set of states and of the instability of the planar spiral
however gave us a hint where to look for the global minimum of the energy.

Consider next the magnetic susceptibility of the equilibrium state.
By virtue of having minimum energy it has a positive static susceptibility
  diverging only at the Goldstone points. We also found
that noncoplanar ordering gives rise to a strong mixture between
in-plane and out-of-plane fluctuations, such that
the Goldstone modes in $\chi^{zz}_q$ at ${\bf q} = \pm {\bar {\bf Q}}$
give rise to a static zero mode in $\chi^{+-}_q$ at $(\pi,\pi)$.
 Another zero mode in $\chi^{+-}(q)$ corresponds to the spin rotation
about the $Z$ axis, and appears at ${\bf q}=- {\bar {\bf Q}}$, as in the spiral
case.
Overall, the in-plane
static susceptibility $\chi^{+-}$ has two poles
\begin{equation}
\chi^{+-}({\bf q}) \approx \frac{\chi_\pi}{({\bf q}-\vec \pi)^2}+
 \frac{\chi_{\bar Q}}{({\bf q} + {\bar {\bf Q}})^2}~,
\end{equation}
where the residues $\chi_\pi$ and $\chi_{\bar Q}$ are proportional to
$\Delta_\perp^2$ and $\Delta_\parallel^2$ respectively. Because
$\Delta_\parallel \sim \sqrt{x}$, the
residue of the pole at the incommensurate wave vector $q=-{\bar {\bf Q}}$ is
proportional to the hole concentration,
and is suppressed with respect to the pole at the commensurate wave
vector $(\pi,\pi)$.
We also considered dynamical susceptibility and found that
the spin wave velocity in the
$zz$ channel does not contain any smallness in $x$.
However, when $z \approx 2$, it behaves as $c \sim ((2/z)-1)$.

To summarize, we studied in this paper
the magnetic phases of the 2D Hubbard model at low
doping. We found that the planar spiral $(\pi,Q)$ phase has lower energy
than the N\'{e}el and $(Q,Q)$ phases in some range of a phenomenological
parameter $z$. However, this state has a negative static
susceptibility in a region of $q$ space around $(\pi,\pi)$,
 and is therefore unstable.
We searched for the stable state, and found that it corresponds
to a very different noncoplanar spin configuration. This, in turn,
gives rise to a novel scenario of the  transformation of the equilibrium
magnetic state with doping. Namely, upon doping,
the initial commensurate antiferromagnetic ordering
 remains unchanged  except for a small
decrease in the amplitude of the order parameter. At the same time, doping
gives rise to a
transverse component of the order parameter which forms a spiral in the
plane perpendicular to the direction of commensurate order. This transverse
component  is small to the extent of $x$, and the low-$T$ behavior
at finite doping remains nearly the same as in the commensurate
antiferromagnet~\cite{commensur}.

Finally, we discuss how these results can be applied to the experimentally
studied disordered regions of cuprate superconductors. The hope here is
that the strongest divergence in the susceptibility in the ordered state
corresponds to the maximum value of the peak in $\chi$ in the region
where the correlation length
is finite. In the absence of a long-range order, the actual susceptibility
is a mixture of $\chi^{+-}$ and $\chi^{zz}$.
In the spiral state, both $\chi^{+-}$ and $\chi^{zz}$ had peaks at the
incommensurate momenta $\pm {\bar {\bf Q}}$.
 In the noncoplanar state however, the
peak in $\chi$  at the commensurate $(\pi,\pi)$ point is much stronger
 than at the incommensurate momenta. We thus observe
that although incommensurability in the nonplanar state is much larger
than in the spiral state (inverse pitch of the spiral scales as $\sqrt{x}$),
 experimentally it is very difficult to
distinguish the noncoplanar state from the N\'{e}el state. The noncoplanar
state therefore should yield the same $T-$dependence of spin-lattice relaxation
rate as the $(\pi,\pi)$ phase. This is consistent with the low-frequency NMR
experiments~\cite{nmr}. The situation at higher frequences (where
neutron-scattering experiments were done~\cite{incommen}) is less clear
as high-frequency bosonic
excitations in the spiral phase are positive, and this phase may be
advantageous over noncoplanar state. The possibility of the
frequency crossover between the two phases is now under study.
 Note in passing
that for $U \sim 6t$ relevant to experiments, we obtain $q \approx 6x$ for the
planar spiral state
in a striking agreement with neutron scattering
experiments on $La_{2-x}Sr_xCuO_4$ ~\cite{incommen} which
yield $q/\pi \approx 2x$.

It is our pleasure to thank  A.Abanov, V. Barzykin,
R.Joynt, D. Pines, S. Sachdev and
A. Sokol for useful
discussions. The work was supported by the University of Wisconsin-Madison
Graduate School and Electric Power Research Institute.

\begin{figure}
\caption{a) Spin configuration of a noncoplanar state. Arrows with thick
ends point out of the plane, while those with thick tails --- into the plane.
This configuration is different from the double spiral considered in
\protect\cite{lee}b. ~b) Two adjacent spins in the equilibrium configuration.
 The in-plane
component, $S_\perp \sim x^{1/2}$,
  is small compared to the off-plane component, $S_\parallel$.}
\label{nonplanar}
\end{figure}


\begin{references}
\bibitem{comm2} In fact, there is a small discontinuity ($\sim x^3$)
in the ground state energy at $\Delta_{\perp} \to 0$
because at any finite $\Delta_{\perp}$ the value of
the inverse pitch of a spiral comes from the compatibility of the two
self-consistency equations rather than from the minimization of energy.
This complication is irrelevant to our analysis as the true ground
state has the energy difference with the planar spiral state
much larger than the jump at  $\Delta_{\perp} \to 0$.
\end{references}
\end{document}